%% file: data-driven-observing-scenarios-paper.tex
\documentclass[twocolumn,times]{aastex631}
\usepackage[nolist]{acronym}
\input{acronyms}

\received{2021 August 16}
\revised{2021 September 27}
\accepted{2021 October 8}

\defcitealias{2019PhRvX...9c1040A}{GWTC-1}
\defcitealias{2021PhRvX..11b1053A}{GWTC-2}
\defcitealias{2020LRR....23....3A}{LRR}
\defcitealias{LIGO-T2000012}{LIGO-T2000012}
\defcitealias{public-alerts-user-guide}{User Guide}

\shorttitle{Data-driven observing scenarios}
\shortauthors{Petrov, Singer et al.}

\begin{document}

\title{Data-driven expectations for electromagnetic counterpart searches based on LIGO/Virgo public alerts}

\correspondingauthor{Leo P. Singer}
\email{leo.p.singer@nasa.gov}

\author[0000-0001-5681-4319]{Polina Petrov}
\affiliation{Department of Physics, Carnegie Mellon University, 5000 Forbes Avenue, Pittsburgh, PA 15213, USA}
\affiliation{Department of Physics \& Astronomy, Vanderbilt University,
2301 Vanderbilt Place, Nashville, TN 37235, USA}

\author[0000-0001-9898-5597]{Leo P. Singer}
\affiliation{Astroparticle Physics Laboratory, NASA Goddard Space Flight Center, Mail Code 661, Greenbelt, MD 20771, USA}

\author[0000-0002-8262-2924]{Michael W. Coughlin}
\affiliation{School of Physics and Astronomy, University of Minnesota, Minneapolis, MN 55455, USA}

\author[0000-0002-8359-9762]{Vishwesh Kumar}
\affiliation{Department of Physics, American University of Sharjah, PO Box 26666, Sharjah, UAE}

\author[0000-0002-4694-7123]{Mouza Almualla}
\affiliation{Department of Physics, American University of Sharjah, PO Box 26666, Sharjah, UAE}

\author[0000-0003-3768-7515]{Shreya Anand}
\affiliation{Division of Physics, Mathematics, and Astronomy, California Institute of Technology, Pasadena, CA 91125, USA}

\author[0000-0002-8255-5127]{Mattia Bulla}
\affiliation{The Oskar Klein Centre, Department of Astronomy, Stockholm University, AlbaNova, SE-106 91 Stockholm, Sweden}

\author[0000-0003-2374-307X]{Tim Dietrich}
\affiliation{Institut f\"{u}r Physik und Astronomie, Universit\"{a}t Potsdam, 
Haus 28, Karl-Liebknecht-Str. 24/25, D-14476, Potsdam, Germany}
\affiliation{Max Planck Institute for Gravitational Physics (Albert Einstein Institute),
Am M\"{u}hlenberg 1, Potsdam D-14476, Germany}

\author[0000-0003-4617-4738]{Francois Foucart}
\affiliation{Department of Physics \& Astronomy, University of New Hampshire, 9 Library Way, Durham NH 03824, USA}

\author[0000-0003-1585-8205]{Nidhal Guessoum}
\affiliation{Department of Physics, American University of Sharjah, PO Box 26666, Sharjah, UAE}

\begin{abstract}
Searches for electromagnetic counterparts of gravitational-wave signals have redoubled since the first detection in 2017 of a binary neutron star merger with a gamma-ray burst, optical/infrared kilonova, and panchromatic afterglow.
Yet, one LIGO/Virgo observing run later, there has not yet been a second, secure identification of an electromagnetic counterpart.
This is not surprising given that the localization uncertainties of events in LIGO and Virgo's third observing run, O3, were much larger than predicted.
We explain this by showing that improvements in data analysis that now allow LIGO/Virgo to detect weaker and hence more poorly localized events have increased the overall number of detections, of which well-localized, \emph{gold-plated} events make up a smaller proportion overall.
We present simulations of the next two LIGO/Virgo/KAGRA observing runs, O4 and O5, that are grounded in the statistics of O3 public alerts.
To illustrate the significant impact that the updated predictions can have, we study the follow-up strategy for the Zwicky Transient Facility.
Realistic and timely forecasting of gravitational-wave localization accuracy is paramount given the large commitments of telescope time and the need to prioritize which events are followed up. 
We include a data release of our simulated localizations as a public proposal planning resource for astronomers.
\end{abstract}

\keywords{%
Astronomical simulations (1857),
Gravitational wave astronomy (675),
Optical observatories (1170),
Neutron stars (1108),
Stellar mass black holes (1611)
}

\section{Introduction}
\label{sec:introduction}

The detection of the first \ac{BNS} merger GW170187 \citep{2017PhRvL.119p1101A} by the Advanced Laser Interferometer Gravitational-wave Observatory (\acsu{LIGO}; \citealt{2015CQGra..32g4001L}) and Virgo \citep{2015CQGra..32b4001A}, its short \ac{GRB} 170817A \citep{2017ApJ...848L..14G}, its afterglow \citep[e.g.][]{2017Sci...358.1579H,2017Natur.551...71T}, and its \ac{KN} AT2017gfo \citep[e.g.][]{2017Sci...358.1565E,2017Sci...358.1559K,2017Sci...358.1583K,2017Natur.551...67P,2017Sci...358.1574S,2017Natur.551...75S} has shown significant promise for multimessenger constraints on many areas of physics, including the \ac{NS} equation of state \citep[e.g.][]{2017ApJ...850L..34B,2018PhRvL.121p1101A,2018ApJ...852L..29R,2019MNRAS.489L..91C,2018MNRAS.480.3871C,2020MNRAS.492..863C,2020Sci...370.1450D}, cosmology \citep[e.g.][]{2017Natur.551...85A,2019NatAs...3..940H,2020Sci...370.1450D}, and nucleosynthesis \citep[e.g.][]{2017ApJ...848L..19C,2017Sci...358.1556C,2017ApJ...848L..17C,2017Natur.551...67P}.

By the end of \ac{O2}, GW170187 was merely one of 11 \ac{GW} events reported in the First \ac{GW} Transient Catalog (GWTC-1; \citealt{2019PhRvX...9c1040A}). The tally has already climbed to many tens of events from just the first half of O3 in GWTC-2 \citep{2021PhRvX..11b1053A}. O3 saw the detection of a second \ac{BNS} merger \citep{2020ApJ...892L...3A} and the first two \aclp{NSBH} (\acused{NSBH}\acsp{NSBH}; \citealt{2021ApJ...915L...5A}). These events and others were followed up by many teams \citep[e.g.][]{2019ApJ...885L..19C,2020MNRAS.497.5518A,2020MNRAS.497..726G,2021NatAs...5...46A}. But while hopes ran high in O3 after the success of GW170817, no further \ac{EM} counterparts have yet been confirmed \citep[although see][]{2020PhRvL.124y1102G,2020AstL...45..710P}.

One possible explanation is that the \ac{KN} AT2017gfo, like many objects in astronomy that are the first of their kind, may have been anomalously bright compared to the rest of its class. However, aggregate analysis of many \ac{KN} searches (with one detection and several upper limits) does not currently require AT2017gfo to have been an outlier \citep{2020MNRAS.497.5518A,2020MNRAS.497..726G,2020ApJ...905..145K}. Rather, the lack of \ac{KN} detections in O3 can mostly be explained by the \ac{GW} localization uncertainties in O3, which were much larger than predicted.

Accurate forecasting of the sensitivity, detection rate, and localization accuracy of the global \ac{GW} detector network based on commissioning and observing scenarios is vitally important to the astronomy community for planning observing campaigns, requesting time allocations on existing facilities, and designing and building new telescopes and space missions. Early estimates of \ac{GW} localization performance for \acp{CBC} were based on analytical analysis of the uncertainty in triangulation from time delay on arrival \citep[e.g.][]{2011CQGra..28j5021F}. Leading up to Advanced LIGO and Virgo, there were more realistic forecasts based on analyzing simulated \ac{CBC} signals using the full end-to-end real-time detection and coherent localization software stack \citep{2014ApJ...795..105S} and realistic noise \citep{2015ApJ...804..114B}. In recent years, LIGO, Virgo, and the \ac{KAGRA} have maintained an official Living Reviews in Relativity (\citealt{2020LRR....23....3A}; henceforth \citetalias{2020LRR....23....3A}) describing their commissioning and observing schedules and expected localization precision. \citetalias{2020LRR....23....3A} had predicted a median localization precision in O3 of $\sim$300\,deg$^2$, but the actual median during O3 was an order of magnitude larger, $\sim$2000\,deg$^2$. Why?

In this paper, we show that the discrepancy between the predicted and \emph{as-built} localization performance in O3 was largely due to differences between the \ac{SNR} threshold for detection that was assumed in \citetalias{2020LRR....23....3A} versus what was used in practice. In past observing runs, \ac{CBC} searches relied completely on finding coincident triggers in multiple detectors to eliminate false positives due to instrumental and environmental glitches. But improvements in flagging and excision of bad data and the estimation of \acp{FAR} made it possible in O3 to detect \ac{CBC} signals in just a single detector \citep{2017CQGra..34o5007C,2019arXiv190108580S,2020arXiv201015282G,2020ApJ...897..169N}, or with a reduced network \ac{SNR} threshold when multiple detectors were online. These advances had the positive impact of increasing LIGO and Virgo's astrophysical reach in O3 and increasing the number of candidates detected. However, they had relatively little effect on the rate of detection of nearby, well-localized, \emph{gold-plated} events, which thereby made up a much smaller fraction of events than predicted in \citetalias{2020LRR....23....3A}.

Because \citetalias{2020LRR....23....3A} is used by astronomers planning observing programs, facilities, and missions for follow-up of \ac{GW} events, this inconsistency led to overly optimistic expectations about the sky localization areas and therefore the telescope time required. We present updated observing scenarios similar to \citetalias{2020LRR....23....3A} but using \ac{SNR} thresholds that are more comparable to what was done in practice in O3. We find good agreement with the statistics of public \ac{GW} alerts that were sent during O3. We carry forward the \ac{SNR} thresholds and provide predicted detection and localization performance for the next two observing runs, O4 (expected to begin in mid-2022) and O5 (expected to begin in 2025). We provide a data release of our thousands of simulated detections and sky localizations for O3, O4, and O5, as a public resource to support observing, proposal planning, and mission formulation.

Due to the need to rapidly tile localization regions of up to thousands of square degrees, \ac{KN} searches are largely the province of wide \ac{FOV}, synoptic, time-domain optical survey facilities like the \acl{Pan-STARRS} (\acsu{Pan-STARRS}; \citealt{2012SPIE.8444E..0HM}), the \acl{DECAM} \citep[\acsu{DECAM};][]{2015AJ....150..150F}, \acl{ZTF} \citep[\acsu{ZTF};][]{2019PASP..131a8002B,2019PASP..131g8001G,2019PASP..131a8003M,2020PASP..132c8001D}, and (soon) the \acl{Rubin} \citep[\acsu{Rubin};][]{2019ApJ...873..111I}; or arrays of many smaller-aperture robotic telescopes such as the \acl{GRANDMA} (\acsu{GRANDMA}; \citealt{2020MNRAS.497.5518A}) and the \acl{GOTO} (\acsu{GOTO}; \citealt{2020MNRAS.497..726G}). To illustrate the impact of realistic, data-driven predictions of \ac{GW} localization performance, we simulate observation planning and \ac{KN} detection rates in the context of \ac{ZTF}.

This paper is organized as follows. Section~\ref{sec:simulations} describes the simulations. The simulations are compared with O3 public alerts in Section~\ref{sec:o3comparison}, while the projections are given in Section~\ref{sec:projections}. The effect that the observing scenarios have on wide-field optical surveys are demonstrated in Section~\ref{sec:optical}. Our conclusions are presented in Section~\ref{sec:conclusion}.

\section{Simulation}
\label{sec:simulations}

Our setup and tools are similar to those used for the \ac{CBC} simulations in \citetalias{2020LRR....23....3A}. 
At a high level, we first draw a sample of simulated compact binaries with a realistic astrophysical distributions of masses, spins, distances, and sky locations. Then we simulate their \ac{GW} signals, add them to Gaussian noise, and recover them with a matched filter. Finally, we make sky maps for those events that pass the threshold for detection. Scripts and instructions for reproducing our simulations on an HTCondor or PBS computing cluster are publicly available on GitHub%
\footnote{\url{https://github.com/lpsinger/observing-scenarios-simulations}}
and Zenodo \citep{cluster-scripts}.
All of the simulated detections and localizations have been release publicly on Zenodo \citep{O3-data,O4-data,O5-data,O6-data}.

\paragraph{Analysis tools}
We used the same analysis tools as \citetalias{2020LRR....23....3A}, employing the rapid localization code, BAYESTAR \citep{2016PhRvD..93b4013S}. We ran the same tools to draw sources from the astrophysical population%
\footnote{\url{https://lscsoft.docs.ligo.org/ligo.skymap/tool/bayestar_inject.html}}%
, to simulate the matched filter pipeline and detection thresholds%
\footnote{\url{https://lscsoft.docs.ligo.org/ligo.skymap/tool/bayestar_realize_coincs.html}}%
, to perform sky localization%
\footnote{\url{https://lscsoft.docs.ligo.org/ligo.skymap/tool/bayestar_localize_coincs.html}}%
, and to gather   summary statistics for the sky maps.%
\footnote{\url{https://lscsoft.docs.ligo.org/ligo.skymap/tool/ligo_skymap_stats.html}}

\paragraph{Detector configurations}
We adopt the same \ac{GW} detector network configuration, noise curves, and duty cycles as \citetalias{2020LRR....23....3A}, which described three observing runs, O3, O4, and O5. The O3 run includes the LIGO Hanford, LIGO Livingston, and Virgo, all with their as-built sensitivities. O4 and O5 includes LIGO, Virgo, and KAGRA, with predicted sensitivities based on planned future upgrades and commissioning. Although all three observing runs were included in Figures 1 and 2 of \citetalias{2020LRR....23....3A}, only O3 and O4 were included in detection \deleted{rate} and localization \deleted{accuracy} simulations. We include results for all three observing runs.%
\footnote{At the request of collaborators who are using our simulations as the basis for mission concept studies, we generated an ``O6" configuration that is like O5 but includes LIGO-India as a fifth detector that is at the same sensitivity as the LIGO sites. The O6 simulations are available in Zenodo \citep{O6-data} but are not discussed in this paper.}
We use the same noise \ac{PSD} data files as \citetalias{2020LRR....23....3A}, which were released in \citet{LIGO-T2000012}.%
\footnote{For O3, we used the files aligo\_O3actual\_H1.txt, aligo\_O3actual\_L1.txt, avirgo\_O3actual.txt; for O4, aligo\_O4high.txt, avirgo\_O4high\_NEW.txt, kagra\_80Mpc.txt; for O5, AplusDesign.txt, avirgo\_O5high\_NEW.txt, and kagra\_128Mpc.txt.}
Each detector had a duty cycle of 70\%, uncorrelated with the other detectors.

\paragraph{Mass and spin distributions}
\citetalias{2020LRR....23....3A} employed three separate \ac{GW} source populations: \ac{BNS}, \ac{NSBH}, and \ac{BBH}. \Ac{NS} component masses were normally distributed with mean 1.33\,$M_\sun$ and standard deviation 0.09\,$M_\sun$; \ac{NS} component spin magnitudes were uniformly distributed in $[0, 0.05]$. \Ac{BH} masses were drawn from the interval $[5, 50]$\,$M_\sun$ according to a Salpeter-like power law distribution \citep{1955ApJ...121..161S}, $p(m) \propto m^{-2.3}$; \ac{BH} component spin magnitudes were uniformly distributed in $[0, 0.99]$. Component masses were drawn independently without any constraint on mass ratio. All spins were either aligned or anti-aligned with the orbital angular momentum, with equal probability. Although state-of-the-art population modeling \citep{2021ApJ...913L...7A} now uses more detailed mass and spin distributions than were available at the time of the publication of \citetalias{2020LRR....23....3A}, we adopt the original distributions for ease of comparison.

\paragraph{Spatial distributions}
As in \citetalias{2020LRR....23....3A}, positions and orientations were distributed isotropically. Redshifts were drawn uniformly in co-moving rate density, e.g., $\frac{d}{dz} p(z) \propto \frac{1}{1+z} \frac{dV_C}{dz}$, employing cosmological parameters from \citet{2016A&A...594A..13P}.

\paragraph{Astrophysical rate density}
The astrophysical rate density does not affect the distributions of events, but does affect the predicted detection rate. \citetalias{2020LRR....23....3A} used observational estimates of the \ac{BNS} and \ac{BBH} merger rate densities from \ac{O2} \citep{2019PhRvX...9c1040A,2019ApJ...882L..24A}, and pre-Advanced LIGO constraints on the \ac{NSBH} merger rate density \citep{2010CQGra..27q3001A}. The values in units of Gpc$^{-3}$\,yr$^{-1}$ were 110--3840, 0.6--1000, and 25--109, for \ac{BNS}, \ac{NSBH}, and \ac{BBH}, respectively. Here, we use the latest published observational rate density estimates from O3 \citep{2021ApJ...913L...7A,2021ApJ...915L...5A} of $320^{+490}_{-240}$, $130^{+112}_{-69}$, and $23.9^{+14.3}_{-18.6}$ respectively.

\paragraph{Sample size}
For each combination of detector configuration and source population, we drew $10^6$ simulated mergers, 10 times the sample size used in \citetalias{2020LRR....23....3A}.

\paragraph{Minimum number of detectors}
Our simulations differ from \citetalias{2020LRR....23....3A} in only two important ways: the minimum number of detectors and the minimum \ac{SNR} threshold for detection. \citetalias{2020LRR....23....3A} required a single-detector \ac{SNR}\,$>$\,4 in at least two detectors\deleted{ and a network \ac{SNR} (square root of the sum of the squares of the \acp{SNR} of the individual detectors) $>12$}. This assumption did not accurately represent O3 because some of the \ac{CBC} search pipelines that were participating in public alerts only needed a signal to be present in a single detector \citep{2017CQGra..34o5007C,2019arXiv190108580S,2020arXiv201015282G,2020ApJ...897..169N}. Indeed, GWTC-2 \citep{2021PhRvX..11b1053A} lists four public alerts in the first half of O3 (S190718y, S190901ap, S190910h, and S190930t) that were detected in a single detector. For these reasons, in the present work we drop \citetalias{2020LRR....23....3A}'s requirement of a coincidence in two or more detectors.

\paragraph{\ac{SNR} threshold for detection}
\citetalias{2020LRR....23....3A} required a network \ac{SNR} (square root of the sum of the squares of the \acp{SNR} of the individual detectors) $>12$ for detection, but this did not accurately describe O3 either. In reality, the threshold for a public alert is based on \ac{FAR}, which depends on time and data quality in complicated ways that are hard to predict or model. Both \citetalias{2020LRR....23....3A} and this work follow a longstanding convention of using a network \ac{SNR} threshold as a proxy because \ac{FAR} tends to vary monotonically with network \ac{SNR}. But it remains difficult to pin down a value for the network \ac{SNR} threshold from publicly available information because LIGO and Virgo do not release the \acp{SNR} in public alerts. GWTC-2 \citep{2021PhRvX..11b1053A} lists \acp{SNR} for a subset of public alerts that were recovered in offline re-analysis with \acp{FAR}$\leq 2$\,yr$^{-1}$; the lowest was S190426c with a network \ac{SNR} of 10.1. However, there are new events that were first published in \citet{2020LRR....23....3A} that have a network \acp{SNR} as low as 8.0. We varied the network \ac{SNR} threshold in our simulation until we obtained reasonable agreement with the localization area and luminosity distance distributions of O3 alerts; we arrived at a network \ac{SNR}\,$>$\,9 for \acp{BBH} or $>$\,8 for \acp{BNS} and \acp{NSBH}. We confirmed that this was a reasonable description of how the online GstLAL compact binary search pipeline was operated in O3; the threshold was intentionally made looser for \ac{BNS} and \ac{NSBH} than \ac{BBH} in order to favor alerts for binaries that were likely to produce \ac{EM} counterparts (R. Magee, private communication). In short, based on discussions with the search teams and comparison with the public alert distributions, we \deleted{simply} require a network \ac{SNR}\,$>$\,9 for \acp{BBH} or $>$\,8 for \acp{BNS} and \acp{NSBH}.

\section{Comparison with O3 public alerts}
\label{sec:o3comparison}

We check our simulation by comparing the predicted distribution for O3 to the measured empirical distribution of LIGO/Virgo public alerts. The script that we used to query LIGO and Virgo's public \ac{GraceDB} is available on GitHub.%
\footnote{\url{https://github.com/lpsinger/observing-scenarios-simulations/blob/master/get-public-alerts.py}}
All of the retrieved alerts are listed in Table~\ref{tab:alerts}.

\startlongtable
\begin{deluxetable*}{lrrD|rr|@{\extracolsep{4pt}}rrrrr}
\tablecaption{\label{tab:alerts}Public Alerts in O3}
\tablecolumns{12}
\tablewidth{0pt}
\tablehead{
    \colhead{} &
    \colhead{Dist.\tablenotemark{a}} &
    \colhead{Area\tablenotemark{b}} &
    \multicolumn{2}{c}{Vol.\tablenotemark{c}} &
    \multicolumn{2}{c}{Properties (\%)} &
    \multicolumn{5}{c}{Classification (\%)}
    \\
    \cline{6-8}
    \cline{8-12}
    \colhead{Superevent} &
    \colhead{(Mpc)} &
    \colhead{(deg$^2$)} &
    \multicolumn{2}{c}{($10^6$\,Mpc$^3$)} &
    \colhead{NS} &
    \colhead{Rem} &
    \colhead{BNS} &
    \colhead{NSBH} &
    \colhead{BBH} &
    \colhead{MG} &
    \colhead{Terr}
}
\decimals
\startdata
\cutinhead{BNS (8 events)}
 S190425z & 150 & 10000 &      9.7 & 100 & 100 & 100 &  0 & 0 &  0 &  0 \\
 S190426c & 370 &  1300 &       18 & 100 & 100 &  49 & 13 & 0 & 24 & 14 \\
 S190510g & 260 &  3500 &       22 & 100 & 100 &  42 &  0 & 0 &  0 & 58 \\
 S190718y & 190 &  7200 & $\infty$ & 100 & 100 &   2 &  0 & 0 &  0 & 98 \\
S190901ap & 240 & 14000 &       53 & 100 & 100 &  86 &  0 & 0 &  0 & 14 \\
 S190910h & 240 & 24000 &       79 & 100 & 100 &  61 &  0 & 0 &  0 & 39 \\
 S191213g & 190 &  1400 &      2.9 & 100 & 100 &  77 &  0 & 0 &  0 & 23 \\
 S200213t & 220 &  2600 &       13 & 100 & 100 &  63 &  0 & 0 &  0 & 37 \\
\cutinhead{NSBH (7 events)}
S190814bv & 280 &    38 & 0.17 & 100 &   0 & 0 & 100 & 0 &   0 &  0 \\
 S190910d & 600 &  3800 &  170 & 100 &   0 & 0 &  98 & 0 &   0 &  2 \\
 S190923y & 430 &  2100 &   39 & 100 &   0 & 0 &  68 & 0 &   0 & 32 \\
 S190930t & 110 & 24000 &  7.4 & 100 &   0 & 0 &  74 & 0 &   0 & 26 \\
S191205ah & 360 &  6400 &  150 & 100 &   0 & 0 &  93 & 0 &   0 &  7 \\
S200105ae & 260 &  7700 &   33 &  98 &   0 & 0 &   3 & 0 &   0 & 97 \\
 S200115j & 320 &   910 &  9.3 & 100 & 100 & 0 &   0 & 0 & 100 &  0 \\
\cutinhead{BBH (40 events)}
S190408an & 1500 &  390 &  140 &  0 & 12 & 0 & 0 & 100 &   0 &  0 \\
 S190412m &  820 &  160 &   13 &  0 & 12 & 0 & 0 & 100 &   0 &  0 \\
S190421ar & 2300 & 1900 & 1800 &  0 &  0 & 0 & 0 &  97 &   0 &  3 \\
S190503bf &  420 &  450 &  7.7 &  0 &  0 & 0 & 0 &  96 &   3 &  0 \\
S190512at & 1300 &  400 &  120 &  0 &  0 & 0 & 0 &  99 &   0 &  1 \\
S190513bm & 2000 &  690 &  500 &  0 &  0 & 0 & 1 &  94 &   5 &  0 \\
 S190517h & 2800 &  940 & 1900 &  0 &  0 & 0 & 0 &  98 &   2 &  0 \\
S190519bj & 3100 &  970 & 1800 &  0 &  0 & 0 & 0 &  96 &   0 &  4 \\
 S190521g &  660 & 1200 &   61 &  0 &  0 & 0 & 0 &  97 &   0 &  3 \\
 S190521r & 1100 &  490 &   84 &  0 &  0 & 0 & 0 & 100 &   0 &  0 \\
S190602aq &  780 & 1200 &  100 &  0 &  0 & 0 & 0 &  99 &   0 &  1 \\
S190630ag & 1000 & 8500 & 1500 &  0 &  0 & 0 & 1 &  94 &   5 &  0 \\
S190701ah & 1000 &   67 &   11 &  0 &  0 & 0 & 0 &  93 &   0 &  7 \\
S190706ai & 5700 & 1100 & 6300 &  0 &  0 & 0 & 0 &  99 &   0 &  1 \\
 S190707q &  800 & 1400 &  120 &  0 &  0 & 0 & 0 & 100 &   0 &  0 \\
 S190720a & 1100 & 1500 &  250 &  0 &  0 & 0 & 0 &  99 &   0 &  1 \\
 S190727h & 1100 & 1400 &  280 &  0 &  0 & 0 & 0 &  92 &   3 &  5 \\
 S190728q &  790 &  540 &   44 &  0 &  0 & 0 & 0 &  95 &   5 &  0 \\
 S190828j & 1800 &  590 &  330 &  0 &  0 & 0 & 0 & 100 &   0 &  0 \\
 S190828l & 1600 &  950 &  470 &  0 &  0 & 0 & 0 & 100 &   0 &  0 \\
S190915ak & 1500 &  530 &  220 &  0 &  0 & 0 & 0 &  99 &   0 &  1 \\
 S190924h &  510 &  510 &   15 & 30 &  0 & 0 & 0 &   0 & 100 &  0 \\
 S190930s &  740 & 2000 &  170 &  0 &  0 & 0 & 0 &   0 &  95 &  5 \\
 S191105e & 1200 & 1300 &  320 &  0 &  0 & 0 & 0 &  95 &   0 &  5 \\
 S191109d & 1700 & 1500 &  880 &  0 &  0 & 0 & 0 & 100 &   0 &  0 \\
 S191129u &  760 & 1000 &   74 &  0 &  0 & 0 & 0 & 100 &   0 &  0 \\
 S191204r &  680 &  100 &  5.5 &  0 &  0 & 0 & 0 & 100 &   0 &  0 \\
 S191215w & 2200 &  920 &  920 &  0 &  0 & 0 & 0 & 100 &   0 &  0 \\
S191216ap &  320 &  300 &    2 & 19 &  0 & 0 & 0 &  99 &   1 &  0 \\
 S191222n &  860 & 2300 &  230 &  0 &  0 & 0 & 0 & 100 &   0 &  0 \\
 S200112r & 1100 & 6200 & 1300 &  0 &  0 & 0 & 0 & 100 &   0 &  0 \\
 S200128d & 4000 & 2500 & 7600 &  0 &  0 & 0 & 0 &  97 &   0 &  3 \\
 S200129m &  920 &   53 &  4.4 &  0 &  0 & 0 & 0 & 100 &   0 &  0 \\
 S200208q & 2800 & 1100 & 2400 &  0 &  0 & 0 & 0 &  99 &   0 &  1 \\
S200219ac & 1500 & 1300 &  540 &  0 &  0 & 0 & 0 &  96 &   0 &  4 \\
S200224ca & 1600 &   71 &   29 &  0 &  0 & 0 & 0 & 100 &   0 &  0 \\
 S200225q & 1200 &  670 &  150 &  0 &  0 & 0 & 0 &  96 &   0 &  4 \\
 S200302c & 1700 & 6700 & 3800 &  0 &  0 & 0 & 0 &  89 &   0 & 11 \\
S200311bg &  930 &   52 &  5.7 &  0 &  0 & 0 & 0 & 100 &   0 &  0 \\
S200316bj & 1200 & 1100 &  320 &  1 &  0 & 0 & 0 &   0 & 100 &  0 \\
\enddata
\tablenotetext{a}{Mean a posteriori luminosity distance}
\tablenotetext{b}{90\% credible area}
\tablenotetext{c}{90\% credible comoving volume}
\end{deluxetable*}

We selected all superevents (aggregated candidates from different templates and pipelines in a narrow sliding time window) from O3 that were not retracted and for which the preferred event was a \ac{CBC} trigger and not an un-modeled burst trigger. For each event, we took the last BAYESTAR sky map that was sent rather than the last sky map because we are interested in the precision of rapid localization for follow-up purposes, rather than the precision of the final parameter estimation. We took the source classification and properties from the last alert that was sent. See the LIGO/Virgo Public Alerts User Guide (\citealt{public-alerts-user-guide}; henceforth \citetalias{public-alerts-user-guide}) for further explanation of the contents of the public alerts.

During O3, the LIGO/Virgo public alerts communicated the source classification as five numbers representing the probabilities of the source belonging to the following fiducial classes: \ac{BNS} (defined as having two components of masses 1--3\,$M_\sun$), \ac{NSBH} (one component with a mass of 1--3\,$M_\sun$ and the other $>5$\,$M_\sun$), \ac{BBH} (both component masses $>5$\,$M_\sun$), MassGap (one or both components have masses of 3--5\,$M_\sun$), and Terrestrial (non-astrophysical). The alerts also contained two source property probabilities, both conditioned upon the source being astrophysical: HasRemnant, providing the probability that the source contained an \ac{NS} that was tidally disrupted without immediately plunging into a final \ac{BH}; and HasNS, the probability that at least one of the compact objects had a mass $<3$\,$M_\odot$.

Since the four astrophysical O3 alert source classes did not match the three \citetalias{2020LRR....23....3A} simulation populations, we make a best attempt to assign each superevent to an \citetalias{2020LRR....23....3A} source population as follows. If the highest-probability astrophysical class was \ac{BNS}, \ac{NSBH}, or \ac{BBH}, then the superevent was assigned to the population of the same name. If the highest-ranked class was MassGap, then we assigned the superevent to the \ac{NSBH} population if HasNS\,$\geq 0.5$, or \ac{BBH} otherwise. Although the \citetalias{2020LRR....23....3A} source populations do not allow for the possibility of a compact object with a mass of 2--5\,$M_\odot$, this ad hoc prescription assigns GW190814, with component masses of $23.2_{-1.0}^{+1.1}$ and $2.59_{-0.09}^{+0.08}$\,$M_\odot$ \citep{2020ApJ...896L..44A}, to the \ac{NSBH} category.

\begin{figure*}
    \includegraphics[width=\textwidth]{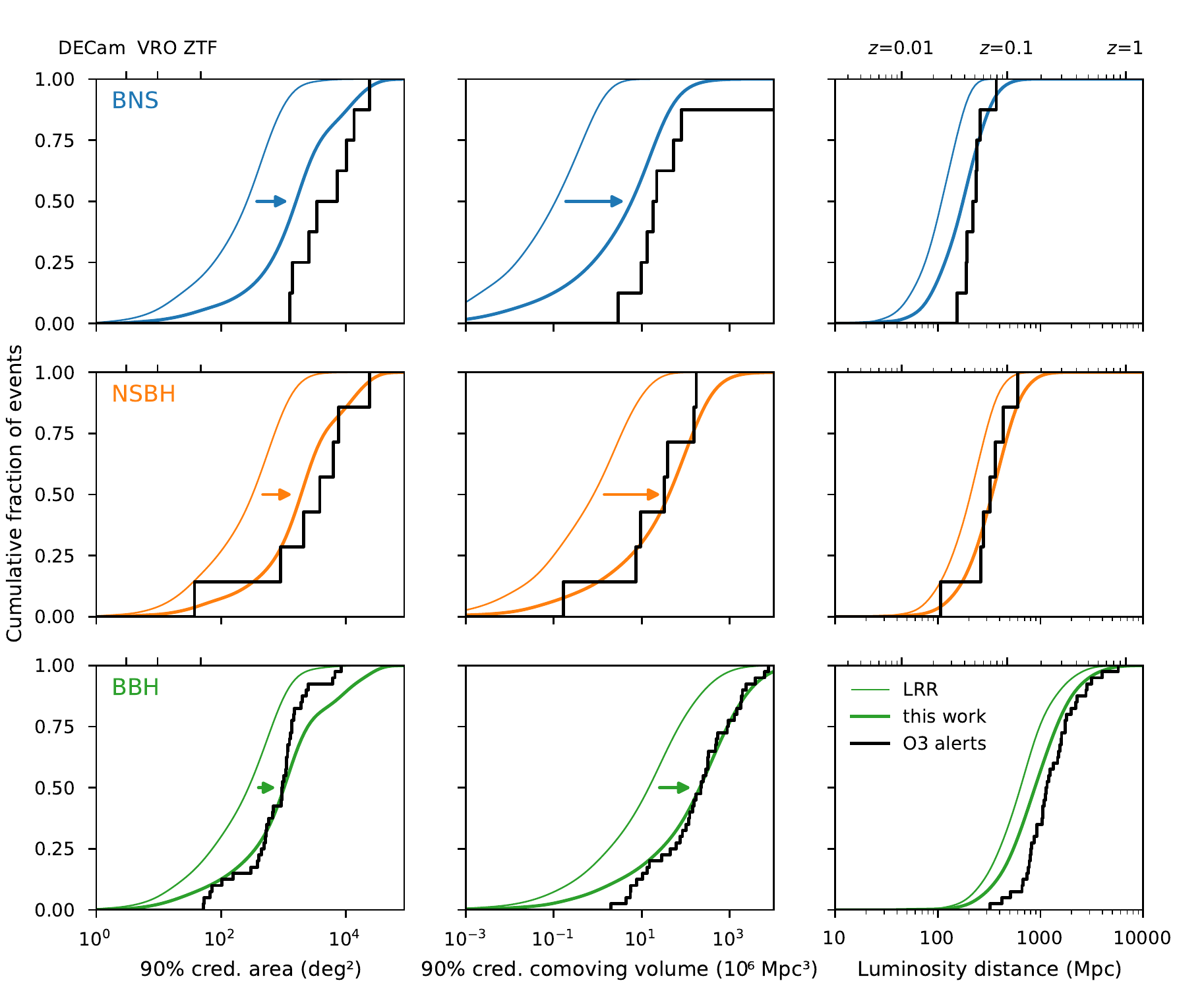}
    \caption{\label{fig:o3-comparison}
    Comparison of simulations with O3 public alerts.
    In each panel, the vertical axis is the cumulative fraction of events.
    Model distributions from \citetalias{2020LRR....23....3A} are shown as thin colored curves and model distributions from this work as thick colored curves.
    The empirical distributions of O3 public alerts are represented by black stepped lines.
    For comparison to modern sky surveys, the \acp{FOV} of \ac{DECAM}, \ac{Rubin}, and \ac{ZTF} are indicated in the left panels.
    }
\end{figure*}

In Fig.~\ref{fig:o3-comparison}, we show the empirical distributions of the 90\% credible areas, 90\% credible co-moving volumes, and estimated luminosity distances of O3 public alerts. We also show the simulated distributions of these quantities from \citetalias{2020LRR....23....3A} and from this work. \citetalias{2020LRR....23....3A} severely underpredicted the distributions of the localization uncertainty and distance provided by the public alerts, whereas the simple change of adjusting the \ac{SNR} thresholds results in much better agreement. For example, \citetalias{2020LRR....23....3A} predicted a median 90\% credible area for \acp{BBH} in O3 of $280^{+30}_{-23}$\,deg$^{2}$, whereas we find $1069^{+43}_{-41}$\,deg$^2$; the sample median of O3 \ac{BBH} alerts was 960\,deg$^2$. The agreement is improved, but not perfect; for example, for \acp{BNS}, \citetalias{2020LRR....23....3A} predicted $270^{+34}_{-20}$\,deg$^2$, whereas we find $1672^{+94}_{-110}$\,deg$^2$, and the sample median was $5400$\,deg$^2$. 

\section{Projections for future observing runs}
\label{sec:projections}

\begin{figure*}
    \includegraphics[width=\textwidth]{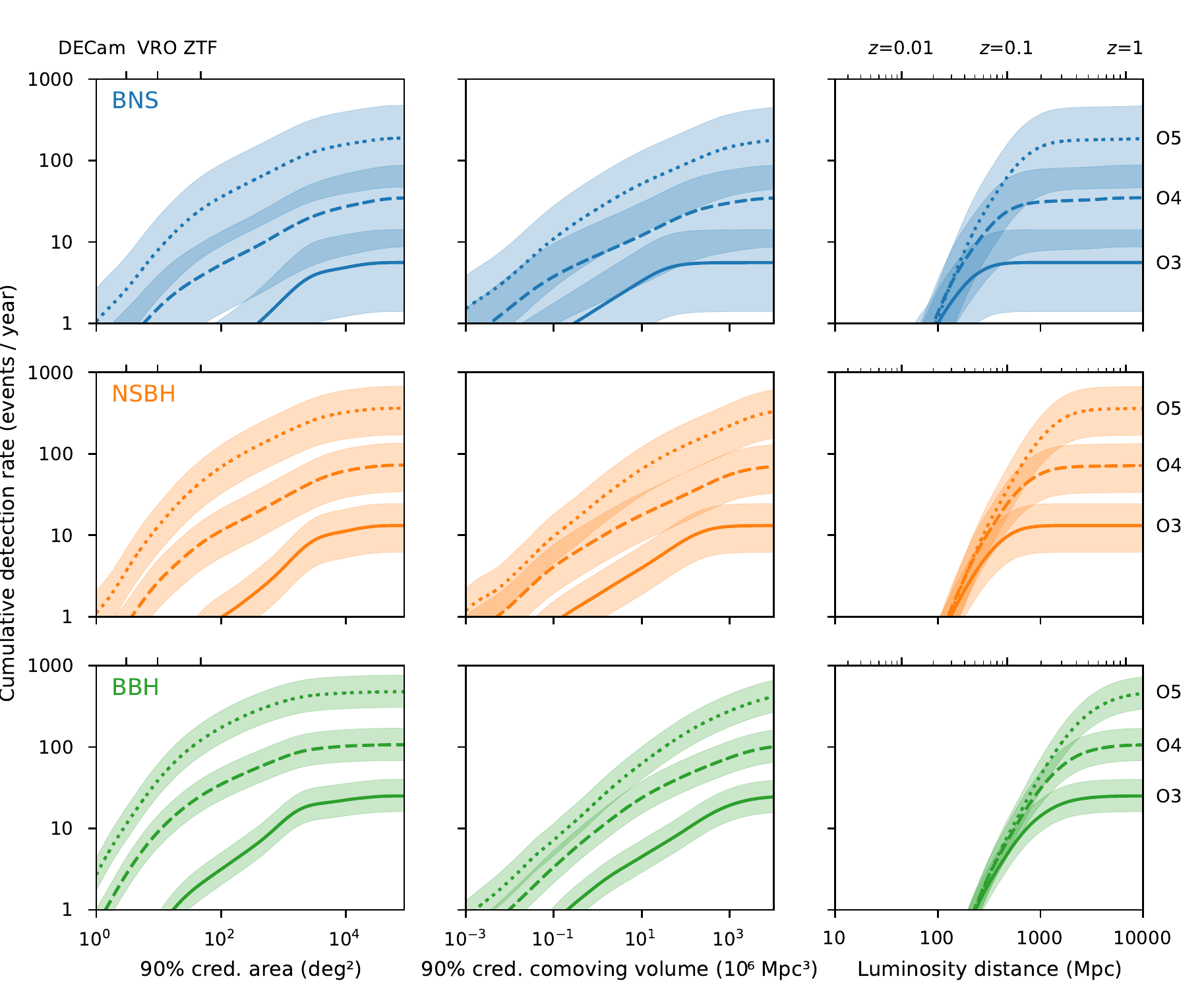}
    \caption{\label{fig:predictions}
    Localization and distance distributions for O3, O4, and O5.
    In each panel, the vertical axis is the cumulative annual detection rate.
    The lightly filled bands around the curves represent the 5--95\% confidence interval due to uncertainty in the astrophysical merger rate density.
    }
\end{figure*}

We now carry forward the new \ac{SNR} threshold in our simulation to provide updated predictions of the detection rate and localization precision for O4 and O5. Fig.~\ref{fig:predictions} shows the cumulative annual detection rate distribution as a function of 90\% credible area, 90\% credible co-moving volume, and distance for these next several observing runs. Shaded bands represent 5\%--95\% variation due to the uncertainty in the astrophysical merger rate density. Table~\ref{tab:summary} presents summary statistics: the median localization area, volume, and distance, as well as the sensitive volume and the annual number of detections. Below, we highlight the most salient features of Fig.~\ref{fig:predictions} and Table~\ref{tab:summary}.

\begin{deluxetable}{cccc}
    \tablecaption{\label{tab:summary}Summary Statistics for O3, O4, and O5.}
    \tablecolumns{4}
    \tablewidth{0pt}
    \tablehead{
        \colhead{Run} &
        \colhead{BNS} &
        \colhead{NSBH} &
        \colhead{BBH}
    }
    \startdata
    \cutinhead{Median 90\% credible area (deg$^2$) \tablenotemark{a}}
    O3  & $1672^{+94}_{-110}$           & $1970^{+110}_{-110}$          & $1069^{+43}_{-41}$        \\
    O4  & $1820^{+190}_{-170}$          & $1840^{+150}_{-150}$          & $335^{+28}_{-17}$         \\
    O5  & $1250^{+120}_{-120}$          & $1076^{+65}_{-75}$            & $230.3^{+7.8}_{-6.4}$     \\
    \cutinhead{Median 90\% credible co-moving volume (10$^6$\,Mpc$^3$) \tablenotemark{a}}
    O3  & $6.62^{+0.97}_{-0.97}$        & $44.1^{+7.4}_{-5.2}$          & $217^{+23}_{-16}$         \\
    O4  & $44.8^{+6.4}_{-6.5}$          & $191^{+20}_{-27}$             & $216^{+16}_{-20}$         \\
    O5  & $125^{+21}_{-12}$             & $448^{+61}_{-44}$             & $538^{+23}_{-24}$         \\
    \cutinhead{Median luminosity distance (Mpc) \tablenotemark{a}}
    O3  & $176.1^{+6.2}_{-5.7}$         & $337.6^{+10.9}_{-9.6}$        & $871^{+31}_{-28}$         \\
    O4  & $352.8^{+10.3}_{-9.8}$        & $621^{+16}_{-14}$             & $1493^{+25}_{-33}$        \\
    O5  & $620^{+16}_{-17}$             & $1132^{+19}_{-23}$            & $2748^{+30}_{-34}$        \\
    \cutinhead{Sensitive volume (10$^6$\,Mpc$^3$) \tablenotemark{a}\tablenotemark{b}}
    O3  & $17.5^{+1.4}_{-1.3}$          & $101.1^{+6.4}_{-6.1}$         & $1047^{+50}_{-49}$        \\
    O4  & $109.0^{+6.7}_{-6.5}$         & $558^{+26}_{-26}$             & $4450^{+130}_{-130}$      \\
    O5  & $590^{+29}_{-28}$             & $2787^{+89}_{-87}$	        & $19950^{+310}_{-310}$     \\
    \cutinhead{Annual number of detections \tablenotemark{c}\tablenotemark{d}}
    O3  & $5^{+14}_{-5}$                & $13^{+15}_{-9}$               & $24^{+18}_{-12}$          \\
    O4  & $34^{+78}_{-25}$              & $72^{+75}_{-38}$              & $106^{+65}_{-42}$         \\
    O5  & $190^{+410}_{-130}$           & $360^{+360}_{-180}$           & $480^{+280}_{-180}$       \\
    \enddata
    \tablecomments{We provide 90\% credible intervals in the form $a^{+b}_{-c}$, where $a$ is the 50th percentile, $a-c$ is the 5th percentile, and $b-a$ is the 95th percentile.}
    \tablenotetext{a}{These credible intervals describe the Monte Carlo sampling uncertainty.}
    \tablenotetext{b}{The sensitive volume is defined as the quotient of the detection rate and the astrophysical merger rate density.}
    \tablenotetext{c}{These credible intervals combine the log-normal uncertainty in the astrophysical merger rate density and the Poisson variation in the number of events over 1\,yr.}
    \tablenotetext{d}{The reader is cautioned of the distinction between the annual detection rate and number of detections in 1\,yr. While the mean number of detections in 1\,yr is equal to the annual detection rate, none of the percentiles of the number of detections scale linearly with the duration of the observing run.}
\end{deluxetable}

\paragraph{The detection rate grows by an order of magnitude in each observing run.}
This is entirely due to the increasing sensitivity of the \ac{GW} detector network. The growth in the detection rate is similar to what was found in \citetalias{2020LRR....23....3A}, although the numbers themselves are a little different due to the updated astrophysical merger rate density estimates and the updated \ac{SNR} threshold. We find a rate of \ac{BNS} detections of 1--14\,yr$^{-1}$ in O3, 9--88 in O4, and 47--478\,yr in O5. This work predicts that the median annual number of \ac{BNS} detections during O5 increases by a factor of $\sim$\,38 from O3, and a factor of $\sim$\,6 from O4. By mid-decade, we predict between tens and hundreds of \ac{BNS} detections per year.

\paragraph{The detection rate of well-localized events grows by an order of magnitude in each observing run.}
The rate of detections of well-localized events, with 90\% credible areas $\leq$100\,deg$^2$, also grows by an order of magnitude in each observing run. For \ac{BNS}, we estimate 0--1\,yr$^{-1}$ in O3 (consistent with the true outcome in O3 of zero alerts fitting these criteria), 1--13\,yr$^{-1}$ in O4, and 9--90\,yr$^{-1}$ in O5. Although events that are localized to $\leq$100\,deg$^2$ constitute a modest fraction of all events, there should be up to tens of them per year by mid-decade.

\paragraph{The median localization area is thousands of square degrees, and does not improve significantly between observing runs.}
This is the most striking impact of the updated \ac{SNR} thresholds. It is plain to see from Figure~\ref{fig:o3-comparison} or by comparing Table~5 from \citetalias{2020LRR....23....3A} with Table~\ref{tab:summary} in the present work. \citetalias{2020LRR....23....3A} predicted median 90\% credible areas for \ac{BNS} events of $270_{-20}^{+34}$\,deg$^2$ in O3 and $33_{-5}^{+5}$\,deg$^2$ in O4, an order of magnitude improvement. We predict a median area of $1672_{-110}^{+94}$\,deg$^2$ in O3, $1820_{-170}^{+190}$\,deg$^2$ in O4, and $1250_{-120}^{+120}$\,deg$^2$ in O5: orders of magnitude worse than \citetalias{2020LRR....23....3A}, and relatively static between observing runs.

\paragraph{The sensitive volume is 3--7 times larger than estimated by \citetalias{2020LRR....23....3A}.}
Table~\ref{tab:summary} also shows the sensitive volume, which is a measure of the space within the universe that is probed by the \ac{GW} detector network, and is defined as the detection rate divided by the astrophysical merger rate density.%
\footnote{We warn the reader that although the 90\% credible co-moving volume and the sensitive volume have the same units, they are different cosmological volume measures. The 90\% credible co-moving volume is defined as an integral over differential co-moving volume. The sensitive volume, on the other hand, has an additional weighting factor of $1/(1 + z)$ to account for time dilation due to the assumption of fixed merger rate density. However, we can compare the values of the 90\% credible co-moving volume and the sensitive volume in Table~\ref{tab:summary} to get a rough sense that \ac{BNS} events are typically localized to 20\%--40\% of the volume probed by the \ac{GW} detector network.}
Although \citetalias{2020LRR....23....3A} itself does not quote values for the sensitive volume, the \citetalias{public-alerts-user-guide} does quote values from the same simulations. Depending on the source class and observing run, we estimate a sensitive volume that is 3--7 times larger. This is partly due to the change in the network \ac{SNR} threshold: if detections were isotropic, we would expect the volume to change by the cube of the ratio of the old and new network \ac{SNR} thresholds, $(12/8)^3 = 3.375$ for \ac{BNS} and \ac{NSBH}, $(12/9)^3 = 2.370$ for \ac{BBH}. A minor fraction of the improvement comes from the increased live time due to the added capability of detecting events when only a single \ac{GW} detector is online: given three detectors with independent duty cycles of 70\%, there is only one detector online $3 \times 70\% \times (30\%)^2 = 18.9\%$ of the time. A more significant fraction of the improvement in sensitive volume comes from the increased isotropy in the sensitivity of the \ac{GW} detector network when two or more detectors are online due to the effective removal of the single-detector \ac{SNR} threshold. For example, a signal that is near the maximum of LIGO Hanford's antenna pattern but near the minimum of Virgo's might register single-detector \acp{SNR} of 12 and 1 respectively, and a network \ac{SNR} of $\sqrt{12^2 + 1^2}$; while \citetalias{2020LRR....23....3A} would not have considered this a detection, the simulations in this work would. Consequently, the reduction in \ac{SNR} thresholds allows the \ac{GW} detector network to probe 3--7 times more of the Local Universe.

\section{Impact on optical searches}
\label{sec:optical}

We now study the impact on prospects for \ac{EM} counterpart searches in O4 and O5. We focus on \ac{ZTF} \citep{2019PASP..131a8002B,2019PASP..131g8001G,2019PASP..131a8003M,2020PASP..132c8001D} because it is especially well adapted to \ac{EM} counterpart searches \citep[e.g.][]{2019ApJ...885L..19C,2021NatAs...5...46A}. ZTF consists of an optical camera with an exceptionally large 47\,deg$^2$ \ac{FOV} on the fully robotic, meter-class \acl{P48}. When interrupted from its preplanned surveys by a \ac{GW} \ac{TOO}, it can rapidly tile large swaths of the sky. The fast 3 day cadence of the \ac{ZTF} Northern Sky Survey provides up-to-date premerger images, aiding in ruling out foreground false positives over the entire accessible sky.

\paragraph{Observing strategy}
\ac{ZTF} typically observes \ac{GW} \acp{TOO} in both the $g$ and $r$ bands when time permits, but for simplicity we simulate only $r$-band observations because the peak absolute magnitude of the \ac{KN} is similar in the two bands. \ac{ZTF} has generally used $\sim$120--300\,s exposures for \acp{TOO} \citep{2020ApJ...905..145K}, reaching a limiting magnitude of $\sim$21.5--22.4. We adopt an exposure time of 300\,s. We require two exposures separated by 30 minutes in order to rule out faint, uncatalogued, moving solar system objects. In order to be \emph{detected}, the apparent magnitude of the \ac{KN} must be above the limiting magnitude in both exposures.

\paragraph{Observation planning}
We generate optimal \ac{ZTF} observing plans for the simulated \ac{GW} events using the same software that is used to plan real \ac{ZTF} \ac{GW} observations, gwemopt \citep{2018MNRAS.478..692C,2019MNRAS.489.5775C}. It optimizes the selection of fields and the allocation of available observing time based on the sensitivity and observing constraints of the telescope, the \ac{GW} sky map, and the expected light curve of the transient.

\paragraph{Light-curve model}
We use \ac{KN} light-curve models generated by the radiative transfer code POSSIS \citep{2019MNRAS.489.5037B} and first presented in \citet{2020Sci...370.1450D}. The model is axisymmetric and has two ejecta components: a lanthanide-rich dynamical component with mass $M_\mathrm{dyn}$ extending above and below the merger plane by half-angle $\Phi$, and a lanthanide-free wind component with mass $M_\mathrm{wind}$ at higher latitudes. We consider \emph{optimistic} and \emph{conservative} model parameters for the \ac{BNS} and \ac{NSBH} mergers that are listed in Table~\ref{tab:kilonova-params}. The optimistic \ac{BNS} parameters are at the high end of plausible \ac{KN} scenarios and more than twice the total best-fit values for AT2017gfo \citep{2020Sci...370.1450D}. Depending on mass and spin, some \ac{NSBH} mergers may not produce any \ac{KN} emission. Of those that do, they are generally expected to produce more dynamical ejecta than \ac{BNS} mergers \citep{2020ApJ...904..155A} and therefore longer-lasting emission.

\begin{deluxetable}{cllll}
    \tablecaption{\label{tab:kilonova-params}\ac{KN} Model Parameters}
    \tablecolumns{5}
    \tablewidth{0pt}
    \tablehead{
        \colhead{} &
        \multicolumn{2}{c}{Optimisic} &
        \multicolumn{2}{c}{Conservative}
        \\
        \cline{2-3}
        \cline{4-5}
        \colhead{} &
        \colhead{BNS} &
        \colhead{NSBH} &
        \colhead{BNS} &
        \colhead{NSBH}
    }
    \startdata
    Dynamical ejecta mass ($M_\odot$) & 0.005 & 0.08 & 0.01 & 0.01 \\
    Wind ejecta mass ($M_\odot$) & 0.11 & 0.09 & 0.01 & 0.01 \\
    Half opening angle & 45\arcdeg & 30\arcdeg & 45\arcdeg & 30\arcdeg \\
    \hline
    Peak $g$-band absolute magnitude & -15.7 & -15.2 & -15.1 & -14.3 \\
    Peak $r$-band absolute magnitude & -16.0 & -15.4 & -15.7 & -14.8 \\
    \enddata
\end{deluxetable}

\begin{figure*}
    \gridline{\fig{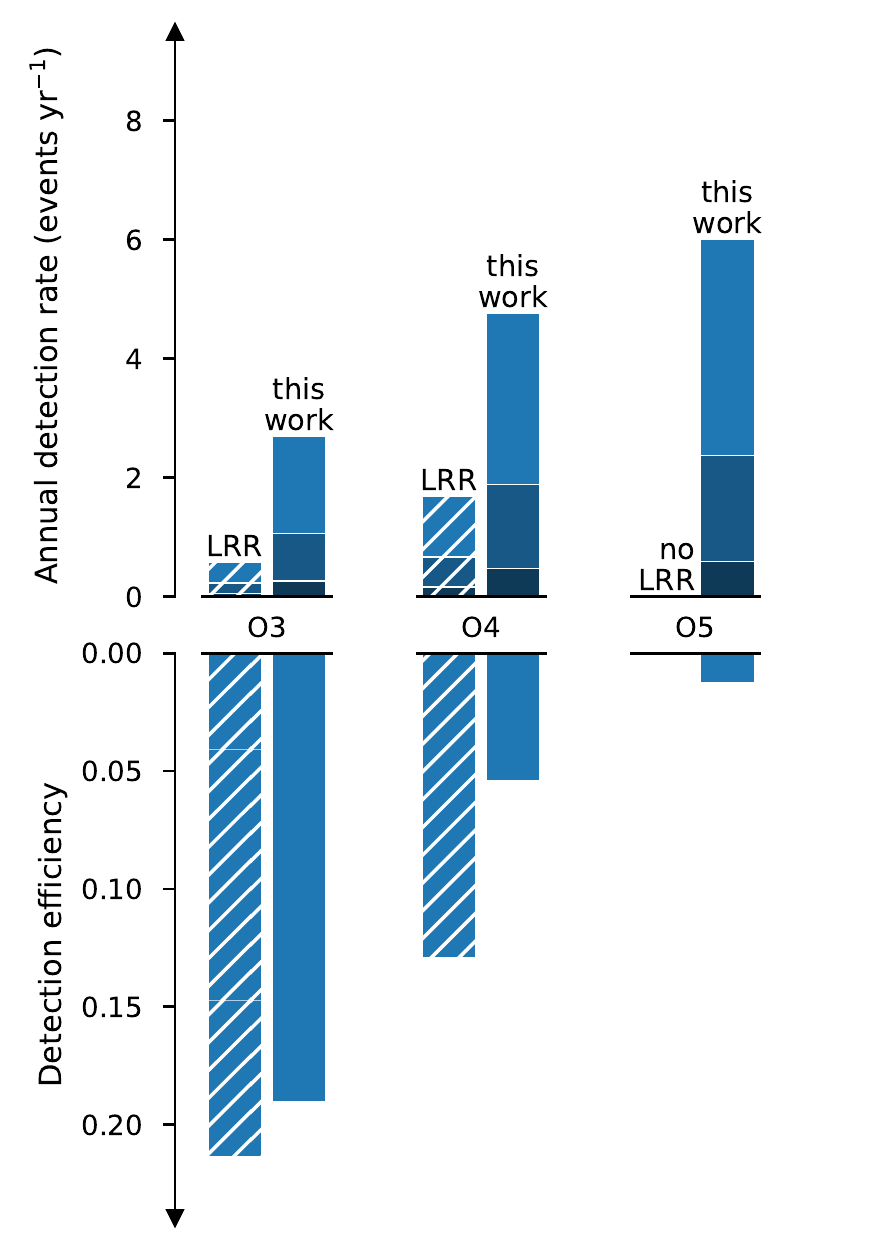}{0.45\textwidth}{(a) Conservative}
              \fig{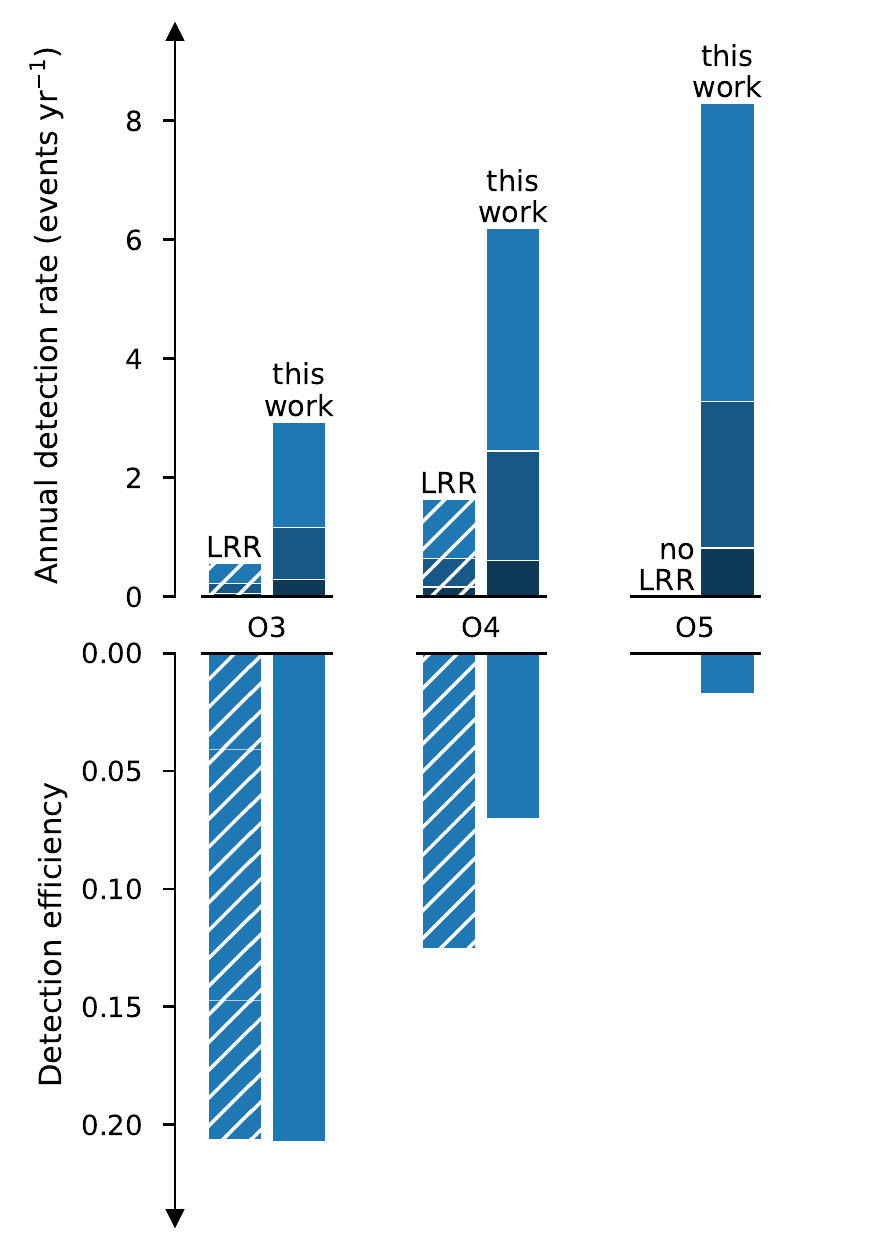}{0.45\textwidth}{(b) Optimistic}}
    \caption{\label{fig:gwemopt}
    Bar chart of predicted \ac{KN} detections of \ac{BNS} events with \ac{ZTF} in O3, O4, and O5. The left panels (a) are for the conservative \ac{KN}, whereas the right panels (b) are for the optimistic case. For comparison, predictions from \citetalias{2020LRR....23....3A} are shown with diagonal hashing. The top panels show the annual detection rate; each bar is further divided into three shaded bands for the 5th, 50th, and 95th percentiles of the astrophysical rate density. The bottom panels show the detection efficiency, defined as the the fraction of \ac{GW}-detected \ac{BNS} events for which \acp{KN} are detected.
    }
\end{figure*}

In Figures~\ref{fig:gwemopt}(a) and Figures~\ref{fig:gwemopt}(b), we present the results of the optical observing simulation under the conservative and optimistic \ac{KN} models, respectively. The top panels show the rate of detectable \acp{KN}, while the bottom panels show the \ac{KN} detection efficiency---the fraction they represent of all detectable \ac{GW} events in the given observing run. We show the outcome under the \ac{SNR} threshold assumptions from both \citetalias{2020LRR....23....3A} and this work while adopting the modern astrophysical rate density estimates for both cases.

We do not show the \ac{NSBH} detection rates and efficiencies in Figure~\ref{fig:gwemopt} because such a small fraction is detected that the Monte Carlo sampling error is very high. Three things work against \acp{NSBH}. First, our \ac{KN} model predicts that \acp{NSBH} are intrinsically 0.5--0.9\,mag fainter in the optical than \acp{BNS}. Second, the \ac{GW} luminosity increases with the binary total mass, leading to a pronounced Malmquist bias \citep{1922MeLuF.100....1M} toward heavier \acp{BH} and more distant \ac{GW} detections. Even with the steeply falling \ac{BH} mass function assumed in \citetalias{2020LRR....23....3A}, \ac{NSBH} detections are on average about twice as far away as \acp{BNS}. Third, if the \ac{BH} is too massive, there should be no \ac{KN} at all because the \ac{NS} must be swallowed by the \ac{BH} before it can be tidally disrupted, although this effect is not captured by our light-curve model. Although these three effects make the prospects for \ac{NSBH} \acp{KN} look bleak, there is room for hope because the two known \ac{NSBH} mergers to date both had relatively low mass \acp{BH} of $8.9_{-1.5}^{+1.2}$ and $5.7_{-2.1}^{1.8}$\,$M_\odot$, respectively.

The main impacts of the more realistic \ac{SNR} thresholds on \ac{KN} searches are higher detection rates but lower detection efficiencies. For example, in O4, the (conservative \ac{KN} model) detection rate nearly triples from 0.2--1.7\,events\,yr$^{-1}$ to 0.5--4.8\,events\,yr$^{-1}$ while the detection efficiency is more than cut half from 13\% to 5\%. Both effects are a direct consequence of the trends that are evident in Figure~\ref{fig:o3-comparison}: the less selective \ac{SNR} threshold for \ac{GW} detection in this work as compared to \citetalias{2020LRR....23....3A} leads to more events overall, but events that are in general more distant, more coarsely localized, and consequently more challenging to follow up. The optimistic \ac{KN} model predicts about 1.3 times as many detections in O4; the detection rate increases to 0.6--6.2\,events\,yr$^{-1}$ and the efficiency increases to 7\%.

Two trends with time are apparent in Figure~\ref{fig:gwemopt}: the predicted \ac{KN} detection rate climbs in each successive observing run while the \ac{KN} detection efficiency drops. We can trace these trends back to the distributions of \ac{GW} events in Figure~\ref{fig:predictions}. The increasing \ac{KN} detection rates are a consequence of an increasing rate of well-localized events, which is a consequence of the improving sensitivities of the \ac{GW} detector---and also a consequence of the addition of a fourth detector, KAGRA. The decreasing \ac{KN} detection efficiencies are a result of the increasing average distances of \ac{GW} detections: with \ac{ZTF}'s limiting magnitude of $r = 22.4$\,mag, it can detect \acp{BNS} out to 417--479\,Mpc. By O5, the median luminosity distance (see Table~\ref{tab:summary}) is greater than that, so \ac{ZTF}'s sensitivity becomes the limiting factor in the \ac{KN} detection efficiency. This highlights the importance with meter-class optical facilities like \ac{ZTF} of being selective about which \ac{GW} events to trigger on. It also draws attention to the importance of future facilities with larger collecting area like \ac{Rubin} to probe more distant \acp{KN} and increase the \ac{KN} sample size.

\section{Conclusion}
\label{sec:conclusion}

In this paper, we have presented simulations of the next two planned \ac{GW} observing runs and their implications for \ac{EM} counterpart searches. We used the public alerts from the third observing run to tune the simulations to be consistent with the observed sky maps, and carried forward the resulting detection thresholds for the next two observing runs. We have used the same source distribution and detector networks as \citetalias{2020LRR....23....3A} to make an \emph{apples-to-apples} comparison with our simulations, quantifying the large impact that the \ac{SNR} threshold assumption has on figures of merit that are important to observers such as detection rate, distance, and sky localization precision. We then performed extensive observational simulations with \ac{ZTF}, to understand the potential multimessenger detection rates for both \ac{BNS} and \ac{NSBH}, with both conservative and optimistic quantities of mass ejecta assumed. Although the updated \ac{SNR} thresholds result in a modest increase in the number of detections, the large distances and large sky areas explain a significant reduction in the fraction of \ac{GW} events for which \acp{KN} are detectable. We demonstrate the need for optical facilities like \ac{ZTF} to adopt selective \ac{TOO} triggering criteria that increase the chances of success. We call attention to the need for deeper observations with upcoming facilities like \ac{Rubin} in order to explore the bulk of the \ac{KN} population, especially with the increasing sensitivity of the \ac{GW} detector network through the middle of this decade.

Our work highlights the need for timely observing scenarios that not only accurately reflect current understanding of the source properties and the detector network configurations, but also the \ac{SNR} thresholds applied to public alerts. We recommend that future revisions of \citetalias{2020LRR....23....3A} use data-driven studies of \ac{GW} alerts from past observing runs to verify that assumptions about \ac{SNR} thresholds lead to realistic results, and promptly incorporate updated thresholds and public alert selection criteria into the official observing scenarios. Because these estimates are relied upon by astronomers to plan observing programs on current facilities, and by observatories and space agencies to plan future instruments and missions for follow-up of \ac{GW} events, it is essential that accurate expectations for detection rates, sky localizations, and distances are provided.

These scenarios also provide quantitative metrics to inform telescope triggering criteria. The rates and simulations encourage targeting smaller localizations from highly significant, nearby events for follow-up with deep observations. Focusing on objects with maximal science return, a fraction will also have \acp{GRB} detected in coincidence; limiting the sample to those with an inclination $\leqslant$10$^{\circ}$ to simulate on-axis events, $\sim$\,5\% of \ac{BNS} from this work can expect a \ac{GRB} detection. Coincidental detection of a \ac{KN} and \ac{GRB} further constrains both the inclination angle and disk mass contribution to the transient, enabling strong \ac{NS} equation of state and Hubble Constant constraints \citep[e.g.][]{2020Sci...370.1450D}.

\begin{acknowledgments}
All of the simulated \ac{GW} localizations and the scripts to reproduce them are publicly available on Zenodo \citep{cluster-scripts,O3-data,O4-data,O5-data,O6-data}.

Resources supporting this work were provided by the NASA High-End Computing (HEC) Program through the NASA Advanced Supercomputing (NAS) Division at Ames Research Center under the project ``Dorado Concept Study Report Monte Carlo Simulations,'' by the Minnesota Supercomputing Institute (MSI) at \ac{UMN} under the project ``Identification of Variable Objects in the Zwicky Transient Facility,'' and the Supercomputing Laboratory at King Abdullah University of Science \& Technology (KAUST) in Thuwal, Saudi Arabia.

P.P. acknowledges support from the \ac{UMN} \ac{REU} sponsored by REU grant No. NSF1757388.
L.P.S. acknowledges support from the Multimessenger Astro Connection Science Task Group (STG) at NASA Goddard Space Flight Center.
M.W.C. acknowledges support from the National Science Foundation with grant Nos. PHY\nobreakdashes-2010970 and OAC-2117997.
S.A. acknowledges support from the National Science Foundation GROWTH PIRE grant No. 1545949.
M.B. acknowledges support from the Swedish Research Council (Reg. no. 2020\nobreakdashes-03330).

This document is LIGO-P2100281-v5.
\end{acknowledgments}

\vspace{5mm}
\facilities{LIGO, EGO:Virgo, Kamioka:KAGRA, PO:1.2m}

\software{%
Astropy \citep{2013A&A...558A..33A},
gwemopt \citep{2018MNRAS.478..692C,2019MNRAS.489.5775C},
HEALPix \citep{2005ApJ...622..759G},
ligo.skymap \citep{2016PhRvD..93b4013S,2016ApJ...829L..15S,2016ApJS..226...10S},
POSSIS \citep{2019MNRAS.489.5037B}.
}

\bibliography{autogenerated-references-do-not-edit-manually,references}{}
\bibliographystyle{aasjournal}

\end{document}

%% file: acronyms.tex
\providecommand{\acrolowercase}[1]{\lowercase{#1}}

\begin{acronym}
\acro{2D}[2D]{two\nobreakdashes-dimensional}
\acro{2+1D}[2+1D]{2+1\nobreakdashes--dimensional}
\acro{2MRS}[2MRS]{2MASS Redshift Survey}
\acro{3D}[3D]{three\nobreakdashes-dimensional}
\acro{2MASS}[2MASS]{Two Micron All Sky Survey}
\acro{AdVirgo}[AdVirgo]{Advanced Virgo}
\acro{AMI}[AMI]{Arcminute Microkelvin Imager}
\acro{AGN}[AGN]{active galactic nucleus}
\acroplural{AGN}[AGN\acrolowercase{s}]{active galactic nuclei}
\acro{aLIGO}[aLIGO]{Advanced \acs{LIGO}}
\acro{ASKAP}[ASKAP]{Australian \acl{SKA} Pathfinder}
\acro{ATCA}[ATCA]{Australia Telescope Compact Array}
\acro{ATLAS}[ATLAS]{Asteroid Terrestrial-impact Last Alert System}
\acro{BAT}[BAT]{Burst Alert Telescope\acroextra{ (instrument on \emph{Swift})}}
\acro{BATSE}[BATSE]{Burst and Transient Source Experiment\acroextra{ (instrument on \acs{CGRO})}}
\acro{BAYESTAR}[BAYESTAR]{BAYESian TriAngulation and Rapid localization}
\acro{BBH}[BBH]{binary black hole}
\acro{BHBH}[BHBH]{\acl{BH}\nobreakdashes--\acl{BH}}
\acro{BH}[BH]{black hole}
\acro{BNS}[BNS]{binary neutron star}
\acro{CARMA}[CARMA]{Combined Array for Research in Millimeter\nobreakdashes-wave Astronomy}
\acro{CASA}[CASA]{Common Astronomy Software Applications}
\acro{CBCG}[CBCG]{Compact Binary Coalescence Galaxy}
\acro{CFH12k}[CFH12k]{Canada--France--Hawaii $12\,288 \times 8\,192$ pixel CCD mosaic\acroextra{ (instrument formerly on the Canada--France--Hawaii Telescope, now on the \ac{P48})}}
\acro{CLU}[CLU]{Census of the Local Universe}
\acro{CRTS}[CRTS]{Catalina Real-time Transient Survey}
\acro{CTIO}[CTIO]{Cerro Tololo Inter-American Observatory}
\acro{CBC}[CBC]{compact binary coalescence}
\acro{CCD}[CCD]{charge coupled device}
\acro{CDF}[CDF]{cumulative distribution function}
\acro{CGRO}[CGRO]{Compton Gamma Ray Observatory}
\acro{CMB}[CMB]{cosmic microwave background}
\acro{CRLB}[CRLB]{Cram\'{e}r\nobreakdashes--Rao lower bound}
\acro{cWB}[\acrolowercase{c}WB]{Coherent WaveBurst}
\acro{DASWG}[DASWG]{Data Analysis Software Working Group}
\acro{DBSP}[DBSP]{Double Spectrograph\acroextra{ (instrument on \acs{P200})}}
\acro{DCT}[DCT]{Discovery Channel Telescope}
\acro{DECAM}[DECam]{Dark Energy Camera\acroextra{ (instrument on the Blanco 4\nobreakdashes-m telescope at \acs{CTIO})}}
\acro{DES}[DES]{Dark Energy Survey}
\acro{DFT}[DFT]{discrete Fourier transform}
\acro{EM}[EM]{electromagnetic}
\acro{ER8}[ER8]{eighth engineering run}
\acro{FD}[FD]{frequency domain}
\acro{FAR}[FAR]{false alarm rate}
\acro{FFT}[FFT]{fast Fourier transform}
\acro{FIR}[FIR]{finite impulse response}
\acro{FITS}[FITS]{Flexible Image Transport System}
\acro{F2}[F2]{FLAMINGOS\nobreakdashes-2}
\acro{FLOPS}[FLOPS]{floating point operations per second}
\acro{FOV}[FOV]{field of view}
\acroplural{FOV}[FOV\acrolowercase{s}]{fields of view}
\acro{FTN}[FTN]{Faulkes Telescope North}
\acro{FWHM}[FWHM]{full width at half-maximum}
\acro{GBM}[GBM]{Gamma-ray Burst Monitor\acroextra{ (instrument on \emph{Fermi})}}
\acro{GCN}[GCN]{Gamma-ray Coordinates Network}
\acro{GLADE}[GLADE]{Galaxy List for the Advanced Detector Era}
\acro{GMOS}[GMOS]{Gemini Multi-object Spectrograph\acroextra{ (instrument on the Gemini telescopes)}}
\acro{GOTO}[GOTO]{Gravitational-wave Optical Transient Observer}
\acro{GRB}[GRB]{gamma-ray burst}
\acro{GraceDB}[GraceDB]{GRAvitational-wave Candidate Event DataBase}
\acro{GRANDMA}[GRANDMA]{Global Rapid Advanced Network Devoted to the Multi-messenger Addicts}
\acro{GROWTH}[GROWTH]{Global Relay of Observatories Watching Transients Happen}
\acro{GSC}[GSC]{Gas Slit Camera}
\acro{GSL}[GSL]{GNU Scientific Library}
\acro{GTC}[GTC]{Gran Telescopio Canarias}
\acro{GW}[GW]{gravitational wave}
\acro{GWGC}[GWGC]{Gravitational Wave Galaxy Catalogue}
\acro{HAWC}[HAWC]{High\nobreakdashes-Altitude Water \v{C}erenkov Gamma\nobreakdashes-Ray Observatory}
\acro{HCT}[HCT]{Himalayan Chandra Telescope}
\acro{HEALPix}[HEALP\acrolowercase{ix}]{Hierarchical Equal Area isoLatitude Pixelization}
\acro{HEASARC}[HEASARC]{High Energy Astrophysics Science Archive Research Center}
\acro{HETE}[HETE]{High Energy Transient Explorer}
\acro{HFOSC}[HFOSC]{Himalaya Faint Object Spectrograph and Camera\acroextra{ (instrument on \acs{HCT})}}
\acro{HMXB}[HMXB]{high\nobreakdashes-mass X\nobreakdashes-ray binary}
\acroplural{HMXB}[HMXB\acrolowercase{s}]{high\nobreakdashes-mass X\nobreakdashes-ray binaries}
\acro{HSC}[HSC]{Hyper Suprime\nobreakdashes-Cam\acroextra{ (instrument on the 8.2\nobreakdashes-m Subaru telescope)}}
\acro{IACT}[IACT]{imaging atmospheric \v{C}erenkov telescope}
\acro{IIR}[IIR]{infinite impulse response}
\acro{IMACS}[IMACS]{Inamori-Magellan Areal Camera \& Spectrograph\acroextra{ (instrument on the Magellan Baade telescope)}}
\acro{IMR}[IMR]{inspiral-merger-ringdown}
\acro{IPAC}[IPAC]{Infrared Processing and Analysis Center}
\acro{IPN}[IPN]{InterPlanetary Network}
\acro{IPTF}[\acrolowercase{i}PTF]{intermediate \acl{PTF}}
\acro{IRAC}[IRAC]{Infrared Array Camera}
\acro{ISM}[ISM]{interstellar medium}
\acro{ISS}[ISS]{International Space Station}
\acro{KAGRA}[KAGRA]{KAmioka GRAvitational\nobreakdashes-wave observatory}
\acro{KDE}[KDE]{kernel density estimator}
\acro{KN}[KN]{kilonova}
\acroplural{KN}[KNe]{kilonovae}
\acro{LAT}[LAT]{Large Area Telescope}
\acro{LCOGT}[LCOGT]{Las Cumbres Observatory Global Telescope}
\acro{LHO}[LHO]{\ac{LIGO} Hanford Observatory}
\acro{LIB}[LIB]{LALInference Burst}
\acro{LIGO}[LIGO]{Laser Interferometer \acs{GW} Observatory}
\acro{llGRB}[\acrolowercase{ll}GRB]{low\nobreakdashes-luminosity \ac{GRB}}
\acro{LLOID}[LLOID]{Low Latency Online Inspiral Detection}
\acro{LLO}[LLO]{\ac{LIGO} Livingston Observatory}
\acro{LMI}[LMI]{Large Monolithic Imager\acroextra{ (instrument on \ac{DCT})}}
\acro{LOFAR}[LOFAR]{Low Frequency Array}
\acro{LOS}[LOS]{line of sight}
\acroplural{LOS}[LOSs]{lines of sight}
\acro{LMC}[LMC]{Large Magellanic Cloud}
\acro{LSB}[LSB]{long, soft burst}
\acro{LSC}[LSC]{\acs{LIGO} Scientific Collaboration}
\acro{LSO}[LSO]{last stable orbit}
\acro{LSST}[LSST]{Large Synoptic Survey Telescope}
\acro{LT}[LT]{Liverpool Telescope}
\acro{LTI}[LTI]{linear time invariant}
\acro{MAP}[MAP]{maximum a posteriori}
\acro{MBTA}[MBTA]{Multi-Band Template Analysis}
\acro{MCMC}[MCMC]{Markov chain Monte Carlo}
\acro{MLE}[MLE]{\ac{ML} estimator}
\acro{ML}[ML]{maximum likelihood}
\acro{MMA}[MMA]{multi-messenger astronomy}
\acro{MOU}[MOU]{memorandum of understanding}
\acroplural{MOU}[MOUs]{memoranda of understanding}
\acro{MWA}[MWA]{Murchison Widefield Array}
\acro{NED}[NED]{NASA/IPAC Extragalactic Database}
\acro{NIR}[NIR]{near infrared}
\acro{NSBH}[NSBH]{neutron star\nobreakdashes--black hole}
\acro{NSBH}[NSBH]{\acl{NS}\nobreakdashes--\acl{BH}}
\acro{NSF}[NSF]{National Science Foundation}
\acro{NSNS}[NSNS]{\acl{NS}\nobreakdashes--\acl{NS}}
\acro{NS}[NS]{neutron star}
\acro{O1}[O1]{\acl{aLIGO}'s first observing run}
\acro{O2}[O2]{\acl{aLIGO}'s and \acl{AdVirgo}'s second observing run}
\acro{O3}[O3]{\acl{aLIGO}'s and \acl{AdVirgo}'s third observing run}
\acro{oLIB}[\acrolowercase{o}LIB]{Omicron+\acl{LIB}}
\acro{OT}[OT]{optical transient}
\acro{P48}[P48]{Palomar 48~inch Oschin telescope}
\acro{P60}[P60]{robotic Palomar 60~inch telescope}
\acro{P200}[P200]{Palomar 200~inch Hale telescope}
\acro{Pan-STARRS}[Pan-STARRS]{Panoramic Survey Telescope and Rapid Response System}
\acro{PC}[PC]{photon counting}
\acro{PESSTO}[PESSTO]{Public ESO Spectroscopic Survey of Transient Objects}
\acro{PSD}[PSD]{power spectral density}
\acro{PSF}[PSF]{point-spread function}
\acro{PS1}[PS1]{Pan\nobreakdashes-STARRS~1}
\acro{PTF}[PTF]{Palomar Transient Factory}
\acro{QUEST}[QUEST]{Quasar Equatorial Survey Team}
\acro{RAPTOR}[RAPTOR]{Rapid Telescopes for Optical Response}
\acro{REU}[REU]{Research Experiences for Undergraduates}
\acro{RMS}[RMS]{root mean square}
\acro{ROTSE}[ROTSE]{Robotic Optical Transient Search}
\acro{S5}[S5]{\ac{LIGO}'s fifth science run}
\acro{S6}[S6]{\ac{LIGO}'s sixth science run}
\acro{SAA}[SAA]{South Atlantic Anomaly}
\acro{SHB}[SHB]{short, hard burst}
\acro{SHGRB}[SHGRB]{short, hard \acl{GRB}}
\acro{SKA}[SKA]{Square Kilometer Array}
\acro{SMT}[SMT]{Slewing Mirror Telescope\acroextra{ (instrument on \acs{UFFO} Pathfinder)}}
\acro{SNR}[S/N]{signal\nobreakdashes-to\nobreakdashes-noise ratio}
\acro{SSC}[SSC]{synchrotron self\nobreakdashes-Compton}
\acro{SDSS}[SDSS]{Sloan Digital Sky Survey}
\acro{SED}[SED]{spectral energy distribution}
\acro{SFR}[SFR]{star formation rate}
\acro{SGRB}[SGRB]{short \acl{GRB}}
\acro{SN}[SN]{supernova}
\acroplural{SN}[SN\acrolowercase{e}]{supernova}
\acro{SNIa}[\acs{SN}\,I\acrolowercase{a}]{Type~Ia \ac{SN}}
\acroplural{SNIa}[\acsp{SN}\,I\acrolowercase{a}]{Type~Ic \acp{SN}}
\acro{SNIcBL}[\acs{SN}\,I\acrolowercase{c}\nobreakdashes-BL]{broad\nobreakdashes-line Type~Ic \ac{SN}}
\acroplural{SNIcBL}[\acsp{SN}\,I\acrolowercase{c}\nobreakdashes-BL]{broad\nobreakdashes-line Type~Ic \acp{SN}}
\acro{SVD}[SVD]{singular value decomposition}
\acro{TAROT}[TAROT]{T\'{e}lescopes \`{a} Action Rapide pour les Objets Transitoires}
\acro{TDOA}[TDOA]{time delay on arrival}
\acroplural{TDOA}[TDOA\acrolowercase{s}]{time delays on arrival}
\acro{TD}[TD]{time domain}
\acro{TOA}[TOA]{time of arrival}
\acroplural{TOA}[TOA\acrolowercase{s}]{times of arrival}
\acro{TOO}[TOO]{target of opportunity}
\acroplural{TOO}[TOO\acrolowercase{s}]{targets of opportunity}
\acro{UFFO}[UFFO]{Ultra Fast Flash Observatory}
\acro{UHE}[UHE]{ultra high energy}
\acro{UMN}[UMN]{University of Minnesota}
\acro{UVOT}[UVOT]{UV/Optical Telescope\acroextra{ (instrument on \emph{Swift})}}
\acro{VHE}[VHE]{very high energy}
\acro{VISTA}[VISTA@ESO]{Visible and Infrared Survey Telescope}
\acro{VLA}[VLA]{Karl G. Jansky Very Large Array}
\acro{VLT}[VLT]{Very Large Telescope}
\acro{Rubin}[Rubin]{Vera C. Rubin Observatory}
\acro{VST}[VST@ESO]{\acs{VLT} Survey Telescope}
\acro{WAM}[WAM]{Wide\nobreakdashes-band All\nobreakdashes-sky Monitor\acroextra{ (instrument on \emph{Suzaku})}}
\acro{WCS}[WCS]{World Coordinate System}
\acro{WSS}[w.s.s.]{wide\nobreakdashes-sense stationary}
\acro{XRF}[XRF]{X\nobreakdashes-ray flash}
\acroplural{XRF}[XRF\acrolowercase{s}]{X\nobreakdashes-ray flashes}
\acro{XRT}[XRT]{X\nobreakdashes-ray Telescope\acroextra{ (instrument on \emph{Swift})}}
\acro{ZTF}[ZTF]{Zwicky Transient Facility}
\end{acronym}